# An Economic Analysis on the Potential and Steady Growth of China: a Practice Based on the Dualistic System Economics in China

By Tianyong Zhou

（National Economic Engineering Laboratory, Dongbei University of Finance and Economics, Beijing, 100096）

**Abstract:**
　　The existing theorization of development economics and transition economics is probably inadequate and perhaps even flawed to accurately explain and analyze a dual economic system such as that in China. China is a country in the transition of dual structure and system. The reform of its economic system has brought off a long period of transformation: "the command system paving the way for the development of a dual system → the coexistence and contentions of the dual system → the integration of the dual system into the market system". From the perspective of economic growth, as the command system gives way to the dual system, the idle and lowly-utilized production factors which have been suppressed during the long period of the command system has begun to be efficiently allocated and utilized, which, in turn, has led to the increased productivity. This is how the economics explains the rapid economic growth potential in the last two decades of the 20th century.

　　The allocation of factors is subjected to the dualistic regulation of planning or administration and market due to the dualistic system, and thus the signal distortion will be a commonly seen existence. Some of the system distortions can be rectified by the economic entities pursuing economic interests, but quite a lot of the system distortions can be irremediable. Compared the systems without distortion, this kind of system distortion causes "low efficiency - idle factors and low utilization of them - institutional factors slack" and output loss in the allocation of production factors. Without the rectification by the continuous market reforms, economic growth will be in a natural stalling state. We have standardized some reference values in the competitive market economy, in order to find out the difference value between various distortion values of dualistic system and those standardized values with the counterfactual method and calculate the new economic growth potential formed by reforms of rectifying system distortion with the logic of market mechanism. There lie two key points--the first is that a classical economic growth model which includes the input of the factor of land must be established seeing that the huge scale of institutional factor slack is caused by the irremediable part without the market-oriented allocation of the land; the second is that the TFP of the countries in the process of institutional transition, unlike the countries with an absolute market economy, mainly comes from the institutional reforms which enhance the factor utilization and improve the factor allocation instead of the technological progress in a broad sense; what is more, the growth of TFP is a curve with small fluctuation or even parallel with the X-axis, which can be observed through the analysis on the data of major countries and regions with a system of market economy.

From the perspective of balanced and safe growth, the institutional distortions of population birth, population flow, land transaction and housing supply, with the changing of export, may cause great influences on the production demand, which includes the iterative contraction of consumption, the increase of export competitive cost, the widening of urban-rural income gap, the transferring of residents' income and the crowding out of consumption. In view of the worldwide shift from a conservative model with more income than expenditure to the debt-based model with more expenditure than income and the need for loose monetary policy, we must explore a basic model that includes variables of debt and land assets that affecting money supply and price changes, especially in China, where the current debt ratio is high and is likely to rise continuously.

Based on such a logical framework of dualistic system economics and its analysis method, a preliminary calculation system is formed through the establishment of models, module coupling and parameters adjusting by programming, to simulate the project arrangement of the institutional reforms of factors focusing on different part or with different extent of forces, the development strategy of water transferring and land improvement and the opening pattern of strategies. In order to predict the economic growth potential, the growth possibility on the side of demand and the security of the debt chain and currency value that can be guaranteed by the assets. All these provides the reference basis of theory, simulation and different results for decision-making in terms.

**A. Introduction:**

Since the late 1970s, various schools of "Transition Economics" have emerged as China, Russia and other countries in the former Eastern Bloc have gradually transitioned their economies from the command economics of state ownership, to the market-oriented economics, the economics of private ownership, or mixed economic systems. The sheer scale of China's transitional economy and the great number of different countries in the Eastern Bloc have inevitably led to divergent value choices, economic theories and practices, which have been grouped into so called "Washington Consensus" and "Beijing Consensus". From the perspective of value orientation, there are two kinds of market economy--one based on private ownership and the other dominated by socialist economy with several types of ownership.

From the perspective of economic methods of reforming, there can be neoclassic economics, Keynesianism featuring the importance of intervention of government, neo-institutional economics and multiple economic schools based on theories of Information Cost, Principal-agent and so on. From the perspective of the practice of institutional transition, we can see a variety of projects and practice including the radical way and progressive way of transition, sequencing approach and the approach of parallel partial progression, and comprehensive and distinctive approach of reform and so on. From the perspective of the transition of the economic system of socialist countries, Russia and eastern European countries have carried out the instantaneous transition to the market economy based on private ownership, which finally comes out with different results. China has launched a progressive transition towards the market economy and the economy dominated by the public ownership with multiple types of ownership, and by 2020, China has become the second largest economic aggregate of in the world with an average annual economic growth rate of 9.2%, and per capita GDP of $10,500.

Unlike most Eastern European countries which have changed its sole form of command economy into sole form of market economy in 5 years or no more than 10 years, China implemented the progressive economic institutional reform which has lasted for over 40 years and is still underway, and such reform may probably go through a vast period spanning from the double increase of GDP since the late 1970s to the preliminary achievement of modernization with a per capita GDP of $20,000 in 2035. That is to say, dualistic economic system transition of China will be a long-term existing status. Such a transition with a large scale, a long spanning of time, a great complexity of practice as well as the profound influence, needs to be taken it as a special researching object, whose process of transition, internal relation, objective law and trend can be revealed and explained by the dualistic institutional economics.

The long-term target of development that per capita GDP of China reaches the level of moderately-developed countries in 2035 has been put forward in the Fourteenth Five-Year Plan, which means the annual GDP growth rate should be kept at 4.73% in 15 years with a stable currency exchange rate of RMB to achieve such goal due to the per capita GDP of moderately-developed countries being between $20,000 to $30,000. Therefore, relevant plans and measures have laid out important tasks in deepening reform in order to seek out the vigor and incentive of the economic development.

The enhancement of reform and institutional transition is more than just a simple transition, which has the ultimate target of solving the low efficiency as well as serious waste under the dualistic system and pursuing the efficient, stable and medium or high speed of growth through the formation of resource allocation in the market economy. Analysis on the basis of Neoclassical Economics on the relationship between economic growth and progressive way of transition reform is still encumbered with the problem of being inadequately applicable. There are some literatures which have measured the quantitative relationship between the promotion of institutional reform and the potential for economic growth, still a logical framework for analyzing China's progressive transition does not seem to have been established.

The basic theoretical research in this field can hardly keep pace with China's complex reform practice, which made it difficult to estimate the importance, and new economic growth potential brought by reform and urgency of reform or to find out the key reform points.

This situation includes three specific aspects:

First, the analysis of the internal relationship between institutional reform and economic growth is confined to be a few suppositions based on comparison and qualitative research. Lawrence and Zheng [7] compared the outputs and growth before and after the reform and found that the reform is likely to generate new outputs. However, this is a quantitative comparison between black-box reform and economic growth, which is not able to analyze the endogenous relationship and structural pattern of institutional reform and economic growth momentum. Such studies cannot provide a reliable and clear basis for government's policy-making on the reform.

Second, no logical relationship between institutional reform and economic growth has been formulated in the neoclassical economic growth model. Since the market economy is exactly the established premise, especially the view that TFP comes from broad technical progress in such model, has led to no inherent logic in the research of some scholars between the analysis paradigm and the meaning of policy, between the input-output analysis and the proposals on reform. As for the researchers and policy-makers, this can be quite misleading since the deepening of education, the progress of knowledge, the accumulation of human capital and the industrialization of new technologies are the most important factors in economic growth instead of the reform in the analysis based on the neoclassical economic growth model where reform as a factor to improve efficiency is ignored.

Third, the factor of land fails to be taken into account in the economic growth model, which means the economic growth potential brought by the market-oriented reform can hardly be reflected in the method. According to the neoclassical economic growth theory, the amount of land available in a country is fixed, and even if the amount of the land for agriculture rises, the proportion of its added value in the total output will continue to decline with its marginal output decreasing. There is also another underlying cause that the land in countries of developed market economy have been capitalized and gone through marketization, therefore, land is not considered to be a variable in the neoclassical economic growth model. Such model still has great defects while being used to analyze the economic growth of a country such as China, where the large-scale land has not been capitalized, utilization level of the land is still low, agricultural land will be converted to urban and non-agricultural land, and the institutional land slack needs to be reinvested when calculating its potential.

The achievement of production capacity needs a corresponding production possibility frontier which has sufficient demand to cover. Since 2012, China's labor force began to diminish iteratively, which will have great impact on the transmission and dynamic trend of the employment-income-consumption chain. Although the declining proportion of export demand to GDP is gradually balanced by the rising investment in real estate, government finance of land and the exorbitantly high housing price still has a diversion and crowding-out effect on income of urban and rural residents and the consumption, which turns out to be the insufficiency of consumption demand as well as the imbalance between supply and demand. This shows the particularity of the situation which combines the strategic transformation of export-oriented industrialization, the delayed influence of population reproduction and the distorted system of land and housing.

The stability of the currency is an important prerequisite of heading towards the level of the developed countries. China has shifted from a country adhering to the value that expenditure depends on the revenue and has a balanced budget with a small surplus, to a country where fiscal deficit, firm and domestic investment under deficit as well as excessive consumption are quite common with a rising debt and more expenditure than income. As far, Gross debt of China has already taken a big proportion of GDP. Slower economic growth rate, low quality of debt and a large shortfall of pension in the future, mean that the debt ratio in the national economy will still increase, and more and more currency will be issued. How to ensure the security of China's debt and monetary system in order to achieve ideal and stable growth is also the important topic to be discussed in mechanism and reform.

This article will focus on the transition theories and practical problems in China, taking attempt to establish a clear logic, a better method and to make a quantitative analysis on the mechanism, in order to build up the analysis method as well as a logic framework which can explain the economic growth miracle of reform and can be applied in the practice of boosting economy at a medium or high rate.

## B. The Sequence of the Market-oriented Reform of Factors and the Mechanism of the Transition of Dualistic Institutional System

Theoretically, the mechanism shown in the history of China's progressive reform is basically the same as it in the sequence of reform of economic factors. From the perspective of dynamic process of the institutional transition, China has been marching on the road of the progressive transition of dualistic economic institution for more than 40 years so far, where 2 essential basic mechanism can be summarized up as follows.

### The Sequence of the Market-oriented Reform of Factors

From the perspective of the sequence logic of product, factor, and marketization reforms, the earliest step is the transformation from planned production of products to market production of commodities; the second step is the reform of the factor allocation system. In which, at first it was free allocation of funds to borrow capitalization, then the reform of labor market-oriented allocation follows, and finally the comprehensive market-oriented allocation reform of land elements is explored and launched. How to promote the capitalization reform of urban and rural resources, production materials, and subsistence materials is still in the stage of discussion, research, and demonstration.

Here, we focus on the sequential logic of the market-oriented reforms of factors. The reform of different factors started with the market-oriented reform of capital that requires a paid use, then marched into the period of the market-oriented reform of labor, and finally arrived the threshold of the market-oriented reform of land, which can be concluded from the sequence of all kinds of reform, the changing of the scale of reform and the development of institution.

The market-oriented reform of capital is a start. Before 1979 state-owned enterprises has a free source of current and fixed assets from the financial allocations from government. Since 1980 some pilot projects of paid occupancy of fixed assets were carried out in some state-owned enterprises with retained profit system. Since 1985, the fixed asset investment of state-owned enterprises has comprehensively changed from financial allocation to loaning, and also the long-term and short-term funds of most self-employed, private and foreign-funded enterprises are allocated by banks and other financial institutions. Later, the long-term investment, which is the source of current assets of enterprises, turned to be supplied by bank loans. In 1990, the Shanghai and Shenzhen Stock Exchanges were established, forming the capital market that included financing of the listed companies, bond issuance and securities transaction and so on. Although there are still monopolies and discrimination in the market of capital factor, the market system of capital allocation regulated by the law of supply -demand and interest is basically formed.

 The market-oriented allocation of labor force originated from the large-scale flow of surplus rural labor force in the late 1980s, and by the beginning of 21st century an early form of market system of labor factor allocation had shown up. Throughout the entire 1980s, urban workers were still recruited by the planned recruitment system, but no system of withdrawing the labor force has been established by any enterprise. There was no contract under the system of market economy between the employers and the laborers. In the late 1980s, China witnessed a large-scale of migrant workers, who mainly worked as supernumerary workers in self-employed, private, foreign-funded or limited liability companies and state-owned enterprises, construction site workers and domestic nannies. Wages were based on the market, or a casual labor contract or oral contract. All these initially led to the formation of a labor market where formal laws or regulations still lacked. Since the 1990s, the country has carried out a series of reforms to promote the market-oriented allocation of labor. It includes the clearing up of the working certificates of migrant workers and various administrative charges, the project of drafting and promulgating the "Labour Law of the People's Republic of China", the reform of laying-off, distributing and reemploying labors in state-owned enterprises and collectively-owned enterprises, the reform of the recruitment in state-owned enterprises, collectively-owned enterprises and public institution which first changed the recruitment from the school allocation system and command system of employers to the two-way choosing between employers and employees, and then developed into graduates' seeking for jobs themselves instead of distributing under the command system, as well as the change from temporary residence permits to residence permits in 2015 for management of the floating population. Since the beginning of the 21st century, the "Labor Contract Law of the People's Republic of China" has been promulgated with the establishment of standards of minimum wage in various places. Also the operation of registered labor-dispatching corporation, where wages were settled down through the negotiation between employers and employees, and multiple networks of labors and specific talents were formed, and headhunters offering the service of finding, introducing and employing the skilled talents of management as well as techniques were allowed.

The market-oriented reform of land took its first step in the 1990s, but due to the unclear goal, distorted system and slow pace of the reform, the allocation of the factor of land still remains the most important part with the greatest force and focus of the planning regulation and administrative control at present. The market-oriented reform of land allocation has just begun, and the construction of unified land market in urban and rural areas may go through a spanning of the next 5-8 years. So the biggest distortion in the economic system lies in the distorted allocation system of factor of land, where there shows the greatest scale of idle or least utilized factor being the institutional land slack. The market-oriented reform on the allocation of the factor of land is of the utmost importance in the future, and the utmost potential of economic growth comes from the betterment of allocation after the market-oriented reform of the factor of both urban and rural land.

**The mechanism in the Process of "Unary Command System-Dualistic System-Unary Market System"**

Based on the view of history, it is the period of transition from unary command system to the dualistic economic system combining both command and market system that comes first, in which more benefits and less cost are brought to people and in which the market is still trying to form its shape. In the second period the enhancement of the dualistic system features where strong relation and severe conflicts between market and command system can be easily seen, and in which comes with the greater difficulty in reform and more tough tasks are to be completed. And the final period is the transition from the dualistic system combining both market and command system to the unary socialist market system, which is the time to make a tough breakthrough of the reform, and in which the hardest part left in the reform, the institution, needs solving.

The period from 1978 to 1998 was the stage of transition from the unary system of command economy to the dualistic system combining the command and market economy in which more economic energy has been exploited. At the early stage of the reform and opening-up, the unary system of command economy which has taken no personal or domestic interests into account has gradually turned into a system of piecework wage, bonus-getting and distribution according to one's performance. More energy economic growth was explored with the expansion of the autonomy of state-owned enterprises and the rise of some collectively-owned enterprises in towns and cities satisfying the demand of horizontal market, in addition to the state-owned and giant collectively-owned enterprises who firmly adhering to the planned economic strategies of supply and selling, indicating the extension of market-oriented economy. A dualistic system of a profit-driven planning coupled with elements of market economy has thus emerged. In 1992, the direction of the reform of socialist market economy was clearly put forward, and the reform of adjusting the structure of ownership, co-existence of various forms of ownership, development of self-employed private economy and introduction of foreign investment was launched. Since the mid-1990s, the ownership reform of state-owned and collectively-owned small and medium-sized enterprises in urban and rural areas has been carried out, focusing more on the larger enterprises and offering more freedom and flexible ways of development for the smaller ones. As a result, the dualistic system where the state-owned and large collectively-owned enterprises coexist with the individual and private and joint-stock economy has expanded, which has lasted for 20 years.

Since 1999, a period of the development and conflict in the dualistic system of market and command economy, where transition has come into a predicament, started.

From the aspect of the allocation of factors, the labor market has been formed with a mobile rural population flow into the cities; while planned administrative controls such as household registration control, education discrimination, the stickiness of land due to its lack of marketization and high cost of housing, have hampered the process of citizenization of peasants.

As for the capital, it can be allocated with a market-oriented system by the signal of supply and demand, interest and stock price while also likely to be distorted by the system in which factors of command economy still exists including monopoly of big banks, loaning and bond-issuing and enterprises' initial public offerings discrimination and so on.

As for the land, on one hand, the existing rural collectively-owned construction land, orchards and farmland of peasants can be transferred in the way of leasing, the housing assets of urban residents can be transacted in the secondary market, and a part of the urban land (such as real estate developers' reserve land and development projects) can be transferred and transacted in the secondary market; on the other hand, there are also rural collective land confiscated by local governments at a low price or monopolized by local governments at a high price, and land planning as well as administrative management such as land-using plan, land-use limit management and prohibition of secondary market transactions of urban and rural land.

Such system with a coexistence of market as well as command economy, and a predicament caused by it, on one hand, has given space of innovation, business-startup, and operation for economic entities; on the other hand, has distorted the markets and blocked the flow of factors. It turns out to be a large amount of idle, least-utilized capital, labor and land factors and the waste of them, from the perspective of supply and demand. Such situation can be called "institutional slack". This period may take 20 years.

From 2021 to 2035, the transition from the dualistic system of command and market economy to the unary socialist market economy should be completed. The task of this stage is to resolutely promote the ultimate transition from the dualistic system to the unary system, rather than rest on, maintain or try to fix the dualistic system.

From the perspective of factor allocation system, what we will see may be as follows:

(1) In terms of population and labor mobility, the birth control of population and household registration control of migration will be abolished, compulsory education will be equalized for new citizens, housing should be supplied via multiple channels including individual building, cooperative building, public rental housing and real estate developers and so on , and the cultivated land and homestead of urbanized population should be retired under the market system by renting, trusteeship, buying shares and selling.

(2) In terms of capital allocation, the degree of concentration and monopoly of the banks should be reduced, the competition in the banking industry will be strengthened, the interest rate between deposits and loans will be narrowed, the discrimination against private enterprises in terms of supply quota of loaning and interest rate should be prohibited, and state-owned enterprises will be held as accountable for the loss of loans as the private enterprises; in terms of bond market, IPO of enterprises, corporate borrowing and IPO of companies, private and state-owned holding enterprises shall be equally accessible, fairly supervised, and the tendency of being biased that more favor comes for the state instead of the public should also be rectified and eliminated.

(3) In terms of land resource allocation, land which belongs to the means of production and living materials has been capitalized at present, the process of assuring all kinds of rights of cultivated land, homestead, collectively-owned construction land and so on should be promoted, the time of the land use will be vastly extended, the rights of all kinds of land will enjoy a free extension when it comes to the 70 years of ownership, property rights of land use can be succeeded, and the land can be leased, invested and mortgaged with an open secondary market for urban and rural lands.

(4) As for the realization form and system-designing arrangement of ownership structure of state-owned enterprises and emerging collectively-owned enterprises, a coupling with operation mechanism of marketization of factors and products to realize a reasonable ownership structure and smooth operation of market system. In the aspect of factor allocation, the socialist unary market economy system will be finally formed.

## C. Non-rectifiable Distortion of Economic System and Institutional Factor Slack

The unary system of command system distorts the signal of horizontal market price and product quantity, and it is a system where only entry, but no transferring or seceding is allowed, which makes low-efficiency producers unable to be retired and high -efficiency producers unable to enter. When the dualistic system where command and market economy coexist, the distortion between the two systems respectively depends on the signal of command administrative signals and market signals will bring more friction and obstacles in the allocation of factors. Whether the thorough distortion of the unary command signal or the distortion of both parts of the dualistic system, will turn out to be different degrees of idling, waste and least utilization of resources in allocation, to be a rising cost of economic operation and a decreasing output level of the economic system. The purpose of reform is to have the distortion in the economic system rectified. The distortion of the system is mainly manifested in the following two aspects: the inability of the smooth operation of coupling between the ownership structure and the market allocation of resources, and the friction between the command adjustment on the factors and the market allocation of factors as well as the loss caused by it. This paper focuses on the dualistic system of allocation of factors, still the importance and key points of the coupling of ownership structure and the operation of the market economy system need a brief explanation herein.

**The Coupling of Land Public Ownership Structure and Market Resource Allocation**

Whether the coupling of land public ownership structure and system of market economy can be a success, is discussed herein from the perspective of the possibility of efficient operation of the mechanism as well as the risk level. Institutional arrangement of the country is needed to be taken into consideration in the dualistic institutional economics. In a socialist market-economy country, the coupling of mechanism like the price depending on the demand and supply and the property right structure of the factors of labor, capital and land, and whether they can work smoothly and efficiently in the whole economic system should be analyzed and solved. When it comes to whether land public ownership and market economy can perfectly fit with each other, following aspects have been focused on in the academia: (1) whether the property rights of the enterprises can enjoy real protection from the law and judicial system; (2)whether the equity of the enterprises can be transferred, transacted and priced besides whether their products can be on sale in the market; (3)whether the property structure and governance structure of public enterprises are equipped with the incentive mechanism, and whether operators are sensitive to and can show a rational attitude towards the economic signals of price, demand and supply, profit and so on; (4) whether there are discrimination on private enterprises from the aspects of market access, the access of factors of all kinds of resource, and tax burden level.

From the perspective of the ownership of factors in China, the factor of labor is individually owned; some of the capital is state-owned, some belongs to the individual and private enterprises, and some is owned by foreign enterprises. Thus, the coupling of ownership structure of factors and market resource allocation of the factor of both labor and capital suffers less friction. Undoubtedly the factor of labor is individually owned, and the fitting and the coupling of multiple ownership of the factor of capital and the market economy is mostly completed. At present, two major problems remain. One is the phenomenon of "Leaving of the Young and Returning of the Old" which means that a surge of the young from the countryside seeking for jobs crowds into the cities while they have to return to the countryside after their Middle Ages caused by system of urban and rural household registration, education, housing and land stickiness, and the other one is the low utilization of capital in the state-owned enterprises.

However, in the socialist market economy with Chinese characteristics, the factor of land, unlike individual ownership of labor and multiple ownership of capital, is owned by the state or collectively owned. The distortion of the allocation of the planned administrative system can be fixed by the flow of labor and capital factors among different economic entities, while the physical attribute of the land that it cannot be easily moved due to its fixed position leads to the inability to rectify the distortion of land allocation caused by the institution by the flowing of factors.

Comprised of broad technical progress, input of factors of labor and capital, and the rate of return, the economic growth model of Solow in the transition economics based on the neoclassical economics does not take the factor of land into account. In countries like China, Vietnam and Laos, which are in the transition to the socialist market economy, A fine design and arrangement of the realization form of ownership to make the market economy and ownership structure organically coupled and operate with low friction and high efficiency, remains to be considered. Because of the ignorance of the factor of land in the neoclassical economics and most of the focus on the capital in the property rights in the institution economics, most of the economists take no account of fitting the ownership of land with the market system, and the necessity of market-oriented reform of the factor of land.

The key to the optimization of the factor allocation through a certain institution is the price and the signal of supply and demand under the market competition besides the interest-oriented economic entities, with a necessity that the factors can be transacted and priced. Labor, capital and land, as the factors of production, are coupled with the operation of market mechanism, and their allocation and transaction should depend on the signals of price as well as the supply and demand. Land can be state-owned or collectively owned or used, but the property rights of land use should be able to transact and to be priced. For instance, labor and capital factors may be owned or used by state-owned or collectively-owned enterprises for a certain period of time, but they must be transacted in the market system when they need to be reallocated. In reality, in the framework of socialist market economy system with Chinese characteristics, the institutional design and arrangement on the factors of labor and capital, which allows the transaction and pricing of both factors, have been accomplished. For enterprises of all types of ownership, the factor of labor is a deal between capital and labor force, with the supply and demand signals of labor market, prices like labor wage and so on. Except a extreme minority of management cadres, the labor force will be transferred and allocated through transactions, even between different but state-owned enterprises. It is impossible not to take into consideration the signals of market supply and demand, not to have price the wages, let alone assign the labor force to another enterprise for free. Capital is deployed between state-owned enterprises through the transaction in the market system. Every capital has an interest price when being offered, state-owned Banks will not provide capital for another state-owned enterprises for free, a state-owned enterprise in the field of metallurgy will never offer capital to the ship-building industry for free just because both are state-owned.

The governmental ownership and collective ownership of land does not affect its market allocation as a factor of production. In the ownership structure of public land, the property right of land use can be transacted and priced, which is a key. Because land is a factor of production, it can and should be allocated by market transactions according to supply, demand, and price signals just like the transfer of the labor and capital factors described above between different firms which however are all state-owned. The nature of land being state-owned and collectively-owned remains unchanged, but its property rights of using should be transacted and allocated by prices based on market supply and demand. The optimization of allocation is needed in the factor of land, just like the factor of labor and capital, and land being able to transact and to be priced is the most basic prerequisite of market economy.

The design of the realization form of land ownership in the field of production and living has been completed in the field of urban housing, but there is still a conflict between the secondary market transaction and the form of property right in the field of urban production. In the form of the realization of ownership, the rural collectively-owned land has not formed the institutional arrangements to fit in and smoothly operate with the socialist market economy. As far as I am concerned, the coupling between the current ownership structure and the market economy operation has not been fully realized, and there are many obstructs, preventing them from smooth operation and causing a very low allocation efficiency of land. From the perspective of conflicts between the current ownership structure and the market economy operation, the utmost importance of property-using rights of land being able to transact and to be priced will be demonstrated herein under the premise that the nature of governmental ownership and collective ownership of land remains unchanged.

First of all, from the perspective of the allocation of urban and rural land, it does not meet with the requirements of the allocation of resources in the socialist market economy. When rural land is being transferred to urban industrial and mining sites, there is no exchange of equal value between two different ownerships, namely the collective ownership and state-owned ownership. The secondary market of urban land has not been opened or even not come into shape due to excessively strict regulations. It is still a mixture of rationing system of command economy and small-scale peasant economy that peasants' homestead is distributed free of charge and the cultivated farmland is contracted under the collective ownership. Land which is collectively-owned, has no clear property rights. Farmers who have to abandon the land for residence and cultivation due to migration fail to do so because the property rights cannot be transacted; farmers want to gather the resource of cultivated land or to purchase homesteads, but eventually find out impossible to realize because the property right cannot be transacted, which shows the factor of land, in fact, is impossible to be allocated in the way of market system. More significantly, the factor of land cannot achieve an optimization of allocation via the market system in the field of rural agriculture, which results in a very low output efficiency and makes the capital unable to join and combine with it. In rural areas, farmers and micro, small and medium-sized enterprises cannot get land for business startup or operation due to land planning and control of use. As a result, the rural secondary and tertiary industries lose the basis for the other factors' combination with land factors to form the output capacity, which profoundly restricts the economic growth capacity in rural areas. Urban industrial and mining enterprises suffers from a lack of secondary transaction market for land, and a great amount of idle and low-utilized land cannot be revitalized through secondary market transaction due to the changing of enterprises, industries, and divisions.

Next comes to the unclear property right of collective ownership of land and the defective structure of property right. From the coupling process of the chain of "property transaction -- output -- distribution -- consumption", this has caused the distortion of distribution and operation including all transaction revenues totally being taken into the local finance, a loss of business startup income based on local features, the widening gap between urban and rural residents' income due to the asset appreciation, the crowded-out consumption because of land finance and high housing prices and so on. While in terms of domestic circulation, this may produce serious consequences of insufficient consumer demand and overproduction. The unequal transaction of land between the urban state-owned economy and the rural collective economy leads to the vast majority of property and land-rent income in the planned allocation of land resources, being either transferred by the local government or privately occupied by the village collective.

Farmers cannot get income from land property and business startup income based on local features, their homestead and cultivated land does not belong to themselves as their assets and wealth, and the formation mechanism of wealth and the process of income distribution are distorted. All of these has widened the gap of income distribution of wealth and assets between rural residents and urban residents (after 1998 houses of urban residents has finally become assets and wealth in terms of market economy after the reform of free distribution turned them into commodity and their monetizing distribution), distorted the process of "income distribution-exchange and expenditures- demand of consumption", hindered the circulation between the "aggregate supply and aggregate demand", and caused a commonly-seen excess capacity and continuing decline of the economic growth rate.

Finally, from the perspective of non-capitalization and non-market-oriented allocation of land, and the security of national economic debt and monetary system, in rural and urban areas especially rural areas, the design of the property right of land has prohibited the land from being allocated by the transaction in the market system, which brings about increasing risks like the breaking of debt chains, sharp fluctuations in currency value, sharp fluctuations of asset price and the plummeting of exchange rates in the tendency of easy monetary policy and debt monetizing. The modern economic model has changed from the mode in the past of pursuing balance of payments and even more left to the mode of pursuing spending more which may cause more debt. In the case that GDP flow cannot balance the expansion of debt and the over issuing of currency, the huge scale of assets of land is the most effective foundation for the stability of the financial and economic system. In the system of ownership structure, the rights of using property cannot be transferred, so that land cannot be transacted, so that it cannot become an asset to be mortgaged to be a guarantee for debt and monetary credit. Stimulating economic growth requires an expansion of debt and money supply, and the scale of assets that can be transacted for mortgage is not sufficient to be an reliable anchor of debt credit and monetary stability. The result of this has to be the breaking of the debt chain, a rocketing asset price and a collapse of the financial system that shares a high possibility.

**Rectifiable and Irremediable Part in the Distortion between the Features of**

**Factors of Production and Allocation of Factors**

The most basic factors of production are labor, capital and land. The dualistic system distorted the market-oriented allocation of factors. However, whether the distortion can be rectified by economic entities in the dualistic system, and how much the irremediable part will influence factors being idle or lowly utilized, do share great relationship with the features of each factor.

Contribution of the factor of labor, to productivity is the wage. After the abolition of the slavery which means the slaves themselves belong to others instead of themselves, the labor forces no more remain to be owned by any private enterprise, or any collectively-owned or state-owned institution, instead they are owned by the individual workers themselves (however, from 1960s to 1980s, the flow of labor factors in China was relatively weak due to the planned recruitment and complicated dismissal process). Because of the characteristics of biological human, labor force is spatially mobile from the side of supply, and workers have their own right to choose and decide whether to be self-employed, or offer labor force in different enterprises, different industries and different regions. As a factor of production and operation, contribution of the factor of capital, to productivity is the profit. It has various form. It may be shown as current assets or fixed assets, physical entities or money and securities. Except buildings, all other capital is spatially mobile, in which currencies and securities has the most liquidity. Contribution of the factor of land, to productivity is the rent. Some land which cannot be separated from buildings is often classified as capital, and the depreciation and profit is the typical manifestation of its rent, but it is spatially immobile. The change of residence and land use can only be realized through the migration of population as well as labor force and the transfer of production. Besides market allocation system, its allocation is often linked with government land plans, controls of use and plans of utilization. The houses that individual's dwell in, although cannot be considered as a factor of production and operation in the micro direct production and operation, they should be taken into account as a factor of productivity in the broad sense, since in the national statistics standard the houses are seen to be paid with a certain rent by the owners in a virtual sense and thus, they will be counted into the GDP.

The reason why we discuss the characteristics of the factors of production is that we need to further analyze the rectifiable and irremediable part of distortion of the economic system of the allocation of factors. In the social production regulation mechanism in the country of the dualistic system, one side is planning and administration, and the other is the market with different levels of competition. Entities in the market system in the dualistic system distortion will try to rectify the distortion of the system through their own behaviors, to pursue their own economic interests. For instance, what labor workers and enterprises react to the distortion of allocation system of factors in the dualistic system like household registration control, discrimination in loaning, land which cannot be transacted and so on, is that the surplus rural labor force temporarily works in cities and towns, state-owned enterprises transfer the funds borrowed from banks to private enterprises, and township enterprises use the land of rural areas (which is prohibited by government regulations) to produce goods to be sold. Thus, migrant workers get their own wages, private enterprises get profits, and the land actually contributes to the land rent. The distortion is rectified, and the loss of output is avoided.

However, to rectify the distortion of factor allocation system, there are premises of necessity and feasibility as follows: (1) an existence of entities pursuing the maximum of economic interests, such as labor workers, entrepreneurs, cooperatives, companies, landlords landowners and so on; (2) all products able to be sold in the competitive market--though homestead cannot be transacted, which leads to the institutional distortion, yet there is an open and competitive market, and the distortion can be rectified by the peasants' attempt to have their service sold in the market through setting up farmland inns and angertainment; (3) a free substitution between factors--for instance, family planning leads to a shortage of labor and rising production costs, and if household registration controls deprive the enterprises of a more stable supply of high-quality labor, then enterprises can use technology like artificial intelligence, online transactions and automated processes and capital to rectify such distortion; (4)the mobility of all factors--for instance, labor force can be set free from the cultivated land which cannot be transacted so that scale production cannot be carried out and the productivity of it appears extremely low, and transferred to the cities to work; (5)an access for economic entities to attain and utilize these factors--for instance, the land in rural areas though is collectively-owned, yet can be attained by distribution and being contracted. Farmers can also produce the products that can be sold in the market on the cultivated land contracted collectively, and enterprises can also build up factories on the land.

However, why further reform of the dualistic system of planning and market is still needed? The reason is that there is institutional distortion of the allocation system of some factors which cannot be rectified, and this will lead to an idling, a waste and a low utilization of factors, with an inevitable output loss.

First is the irremediable distortion in the allocation of factors of capital. Although capital is characterized by strong liquidity, the irremediable distortion of allocation will still exist in the following two situations: (1) If the macro tax burden is too high, the government will distort the allocation of national income between enterprise capital investment and government expenditure, resulting in an output loss with relatively less capital input in the production field. (2) When factors of capital are allocated by the banks and capital market, there will be discrimination between state-owned enterprises and private enterprises, and the principal-agent relationship, insider control and moral hazard transfer of economic entities in state-owned enterprises will also reduce the output efficiency of capital allocation.

After that comes the irremediable distortion in the allocation of factors of labor. The population in the rural agricultural and urban non-agricultural sectors cannot be citizenized, thus which cannot become the stable factor of labor force. Labor workers have their best window period of urbanization in their entire life cycle. However, due to the system distortion of household registration management system, education discrimination, the stickiness of land due to its lack of marketization and high housing cost, the labor process is exactly "Leaving of the Young and Returning of the Old" mentioned above [8]. As a result, many people who have missed urbanization and idle labor forces must be accumulated in the countryside, and the consequence of this distortion is the irreversible process of urbanization. This situation can only be fixed and eased by the new formed output capacity from the combination with institutional land slack, in the countryside, counties, towns, market towns and large villages.

The final part is about the relatively more irremediable distortion in the allocation of factors of land due to all kinds of plans and administrative controls despite the property rights of land being given to the peasants and rural or urban enterprises. In the aspect of the highly concentrated farmland transfer, there are many problems such as high negotiation cost of lease, a common breaking of contracts among peasants, high cost and long time spent in litigation, unstable rent price, agricultural subsidies' failure to reach peasants at last and so on. Therefore, rural families and other rural enterprises, though with the motive to pursue profit maximization, yet cannot utilize these factors of land for the production of various market-oriented products, and thus are unable to rectify and avoid the distortion and output loss which are both caused by the non-market-oriented allocation of land. Agricultural cultivated land cannot avoid contract and price risks through concentrated transaction, and farmers cannot spend their long-term investment in land so that they cannot ideally achieve a long-term stable agricultural scale operation. The same situation is also seen in some land in cities and towns. In the land of state-owned enterprises and public institutions and for national defense, there is a large amount of idle and lowly-utilized land. The transaction in the secondary market of construction land is not allowed, or the transaction is allowed but cannot be finally done in fact because of the excessively long time spent in planning and approval, or the land user cannot change its way of use, all of which will make the distortion in the economic system of the allocation of the factor of land unable to be rectified.

**Market Standard Value vs. System Distortion Value and Institutional Factor Slack**

Is there any analysis way or method in economics, which can be used to calculate the efficiency loss caused by the distortion of the dualistic system, which can estimate how much is the institutional factors slack caused by the difference value from the mathematical aspect, how much efficiency will be improved by the institutional reform, and how much is the potential of its economic growth? In my opinion, when analyzing the country with a long-term transformation of dualistic system, we must theoretically establish a basic concept and the most basic variable of "system distortion value difference". In fact, this phenomenon of system distortion difference value is a common existence in countries with dualistic system, only to be turned a blind eye to by people and scholars who also fail to combine the efficiency, equilibrium, etc. or to think logically about it. What is the system distortion difference value? It is the difference between the relevant standard values in a scenario of the market economy where there is no institutional distortion, and the distortion values caused by the institutional friction in a country with a dualistic system. For example, in 2020, the average rate of return on capital of China's market-oriented private enterprises is about 6%, while the average for state-owned enterprises is less than 2%, then there is a difference value of 4%. Based on the total assets of state-owned enterprises, we can use the difference value to calculate and find out the total output loss of state-owned enterprises, and scale of the idle and lowly-utilized its assets, which together comprise the institutional slack. Its theoretical significance lies in its being an economic variable that describes the most internal connection and essential characteristics of the gap between the results of planned administrative allocation and that of market-oriented allocation, and also being a unique basic economic concept in the countries with dualistic system.

Institutional distortion will increase the friction and barrier between input and output, especially the institutional slack of factors. Therefore, it is necessary to define and explain the institutional slack. Moreover, the stimulation of economic growth potential lies in the utilization of institutional slack by the market reform of factors [9].

In the rural and agricultural fields of the countries with dualistic structural transition, there are many labors forces whose income is lower than that of the urban and industrial fields, and who suffers a low utilization rate or even appears idle, and this is the structural surplus labor force in the developing countries. In the countries in transition, there is in fact a factor surplus caused by institutional distortion, such as the held-up labor forces in the rural areas caused by the institution obstacles of population and labor force migration, excessive personnel in the working positions in urban areas, idle and lowly-utilized land in urban areas and field of production caused by inability to be allocated by market, idle and lowly-utilized fixed assets caused by the economic institutional distortion in state-owned enterprises and so on. Especially when the distortion cannot be rectified by the marketization of products and the behavior of the economic entities to pursue interests, the idle and low utilization of factors caused by the institutional distortion is what we call "Institutional Slack".

Institutional slack is a very important part in mathematical economic analysis of dualistic system transition. If the distortion of allocation system of the factors cannot be rectified, there will be a great number of factors being idle and lowly-utilized, namely, the "Institutional Slack". We can measure the surplus in a variety of ways, either by calculating the output loss caused by institutional distortions through their relevant rents, profits, and wages, or by calculating the amount of output that could be added if the allocation system is reformed and the surplus can be put into production again.

China's structural transformation of the dualistic system is a paradox opposed to Lewis' theories. The institutional slack was first discovered and understood by Tianyong Zhou [10] when he observed data of the complex situation that the shortage of labor and land supply coexists with a large amount of idle labor and land in China's urban and rural areas.

For example, in China, as a developing country, the wage and income gap between urban workers and rural migrant workers, and between urban workers and agricultural peasants has gradually widened. 40 years after 1978, the wage of rural migrant workers in cities has decreased from 150% of that of urban workers in 1978 to 50% in 2019. Peasants in the sector of agriculture have seen their income drop from 50% to 10% of urban wages. In a decade since 2010, the number of migrant workers has decreased from the peak of nearly 20 million to several hundred thousand at the end of this period, but the scale of institutional labor force surplus in rural areas reaches 160 million in 2020. This is obviously not consistent with Lewis' theories [11] of the dualistic transformation model. Originally, structural transformation can gradually reduce the labor force surplus and population surplus in the field of rural and agriculture, but because of the household registration regulation, education discrimination, the stickiness of land due to its lack of marketization and high cost of housing, the one-side way of citizenization of labor force and population has already turned into the phenomenon of "Leaving of the Young and Returning of the Old" mentioned above, which finally leads to a giant amount of institutional labor surplus in rural places. The contradiction between the reality of China and the theories in the Development Economics, can only be explained when it is classified to be the consequence of the institutional distortion.

Take this as another example, on one hand, we have implemented a strict protection system towards cultivated land; on the other hand, the relationship between command allocation of the construction land and the demand of land is still quite severe. In 2020, there will be about 264 million Mu[1] of farmland abandoned or not being taken good care of in rural areas, 86 million Mu of idle land of homestead, and about 40 million Mu of idle or lowly-utilized land in urban areas and industrial as well as mining enterprises. That is to say, when land is only a means of production and living and cannot be allocated by the market, there is also a huge amount of institutional slack.

No matter what kind of economic system in any institution, an average value of all kinds of economic events do exist. The value which the competitive market is assumed to have can be set as a "standard value", and there is also a "distortion value" which can be seen in the dualistic system. It is a crucial part to work out that the difference between the "standard value" and "distortion value" should be observed and calculated, which is shown in Figure 1.

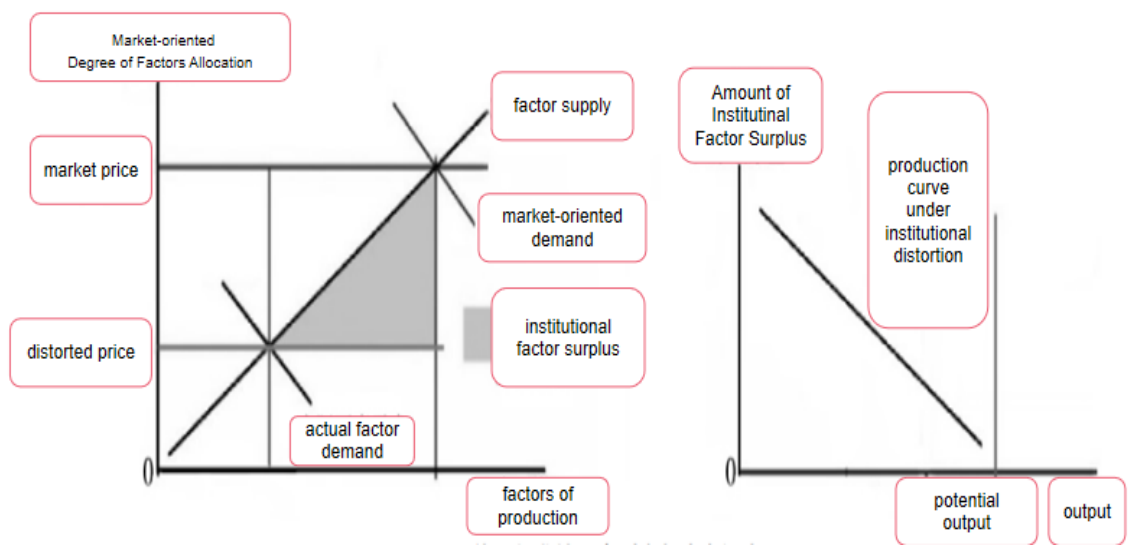

---

[1] Mu, a unit of area (approximately 666.667 squared meters) used in China.

(Figure 1: Institutional Factor Slack under the Distortion and the Output Loss)

As can be seen from Figure 1, in terms of output, under the condition of the same input structure and scale of factors, the input-output level of the competitive market economy is the "standard value", and the input-output level of dualistic system economy is the "distortion value", and the difference between them is the output loss due to the distortion in the dualistic system.

Now let's see such loss from the perspective of input of factors:

In terms of labor mobility, it is the differences between the standard urbanization rate and the distorted urbanization rate, between the standard agricultural employment rate and the distorted agricultural employment rate, between the standard wage level and the distorted wage level when at the same development level; In terms of the factor of capital, it is the differences between the national macro tax burden rate and China's distorted tax burden rate, between the loan and bond interest rate of private enterprises and that of the state-owned enterprises, between the net profit rate of private enterprises and that of the state-owned enterprises. In terms of the factor of land, it is the differences between the land's being able and not being able to be transacted, the price of the private land transaction and the compensation level of the allocated land, the utilization rate of the land in the state of competition and that of the land in the state of distortion. These differences can be measured and observed by comparing horizontally with countries at the same level of development and vertically with countries at the same stage of development, by comparing with countries in the process of transition which are also at the same level of development, by the comparison between the state-owned enterprises which have various operational command indicators and competitive private enterprises, and by comparing the different amount of loans and interest that the capital market and banks offer the enterprises of different ownership.

We can establish the logic based on the counterfactual approach: "what if, in the market allocation, there is not as much as there should be; or more than there should be; even there should be transaction and certain value, but instead there is none, which forms the "distortion of zero". Taking the difference values above as parameters, how much surplus of factors due to system distortion there are and the scale of output loss caused by them were calculated by sampling survey, shadow price and input-output methods.

According to the calculation of the model established in the principle of Figure 1, the situation in 2020 is estimated conservatively as follows: (1) the scale of institutional capital surplus in state-owned enterprises can be 82,701 billion RMB, accounting for 33.83% of the total state-owned assets and 16.95% of the total social capital. (2) the institutional labor force surplus in urban and rural areas is 160 million in rural areas, and 150 million in state-owned administrative enterprises and institutions, which is a total of 175 million people, accounting for 22.86% of the total employed labor force. (3) the country's institutional urban and rural construction land surplus is 151.89 million Mu, accounting for 25.11% of the total urban and rural construction land. Up to 2020, the output loss caused by the idling and low utilization of factors of labor, capital, land and buildings has been 6.30 trillion RMB, 4.92 trillion

RMB and 5.34 trillion RMB respectively. The mismatching between government expenditure and enterprise capital caused by high tax burden also caused 626.6 billion RMB of output loss. The total loss caused by the inefficiency was 17.19 trillion RMB which is 16.92% of GDP of that year. In other words, through the reform to eliminate the inefficiency of the allocation and the low utilization of these factors, there is likely to be a new economic growth potential of 1% approximately on average formed in the next 15 years [12].

D. **TFP Source and Classical Regression Model of Growth in Dualistic System Countries**

From the perspective of a country's national economy, the level of efficiency, whether the national economy is balanced, and whether it can grow safely and steadily are the basic and long-term problems in the operation. However, the distortion of factor allocation caused by the dualistic system leads to low productivity and loss of the output. So, in this case, we will further discuss the relationship between the reform of economic system and the economic growth in countries of the dualistic system.

**The Rate of Natural Economic Increase in the Dualistic System Country in the Process of Transition**

In modern economics, the accurate definition of "the rate of natural economic growth" should be the potential output and growth in a country or region with the system of market economy①[2]. Obviously, for the countries in a long transition, the dualistic system does not meet with this implied condition. So, how to calculate the rate of natural economic growth in the transition countries? In my opinion, the natural economic growth rate of transition countries (that is to say, in the case of no reform, or of an incomplete reform without implementation) should be calculated by using the Solow growth model in the neoclassical economics. The reasons are as follows:

(1) The neoclassical model of output and growth assumes that institution is an exogenous variable that is already set and given, and that it is not distorted. Based on this, we can observe the input-output relationship between technical progress in a broad sense, labor input and capital input with no institutional reform or no implementation of reform. (2) The neoclassical output and growth model does not include the variable of the factor of land. The developed countries have undoubtedly had the land capitalized, and the transition countries also have the reform of land institution not existing or not being implemented, and have the land not capitalized and not ready for redevelopment to improve the utilization rate. Therefore, the variable of the factor of land is allowed to be omitted in the model. (3) TFP is only defined as technical progress in a broad sense, excluding the efficiency improvement caused by the institutional reform and the great reform which boosts the technological progress and technological industrialization, which is in line with the scenario of no reform and no implementation of reform.

---

[2] The rate of natural economic growth, also known as the potential economic growth rate, refers to the growth rate of the maximum amount of goods and services produced by a country or region under the optimal and complete allocation of various resources, or the maximum economic growth rate that can be achieved.

Without finding out the internal logical relationship between the dualistic institutional reform and economic growth, scholars mostly use the most basic analysis tool which is the input-output economic growth model with the structure of three factors of technical progress in a broad sense, labor force and capital. Liu Shijin et al. [13] pointed out that although technical progress in a broad sense played a very important role in China's rapid economic growth, the TFP growth rate has shown a decreasing trend in recent years. Barro [14] points out that China's economy will eventually converge to the same path of world economic growth in the history, and the growth rate may soon fall from 8% to approximately 3% or 4% without any obvious technical progress to pull the future economic growth up. Bai Chongen, Zhang Qiong [15], the research group of Macroeconomic Research Center of Chinese Academy of Social Sciences [16] and Tianyong Zhou [17] point out that, in the absence of obvious technological progress in a broad sense and institutional reform, it has become an inevitable trend that the potential growth of China's economy is slowing down due to the change of population structure. Huang Taiyan and Zhang Zhong [18] believed that in the pessimistic situation that the reform could not achieve the expected effect, China's potential economic growth rate in the period spanning from 2021 to 2025, from 2026 to 2030 and from 2031 to 2035 would be even lower, would be 1.52%, 1.82% and 2.00% respectively.

From the perspectives of input and output, balanced growth and economic security, the team of the National Economic Engineering Laboratory of Dongbei University of Finance and Economics calculates that if there is no more scientific and accurate systematic reform and major development strategic arrangements, and the economy grows in a natural way without any interference no matter what replacement will be between the factors of labor and capital, the average annual GDP growth rate between 2021 and 2035 determined by its input and its output in a broad sense of technological progress is in a range from the top 3.31%, and the bottom 1.81%, within which lies a medium value of 2.5%.

The simulation and characterization of input and economic growth with this analysis tool are consistent with the scenarios in which no major reform is carried out or the major reform is not being implemented. The forecast above of China's future economic growth rate can be regarded as the natural economic growth rate of the countries with a dualistic system.

When the neoclassical economic model of growth is used to analyze the key points of reform and development, it's natural to have its policy meaning assuming that the economic benefits should be improved, which downplays the vital importance of reform in the countries in the process of transition. Since it has the premise that input and output, and the economic growth are carried out in the market system, generally the variables of labor input can be determined, and the variables of capital input can be roughly inferred as well. Whether the economic growth can be stabilized and accelerated is more likely to depend on the TFP, which can be interpreted as the technical progress in a broad sense.

## The TFP of Countries in the Process of Institutional Transition being Mostly from the Reforms and Its Endogenous Calculation

Where does the TFP come from in the model of the output and growth of countries in the process of institutional transition? In fact, under the condition that the amount of the input of factors does not increase, the improvement of utilization rate and the allocation of the factors brought about by vigorous institutional reform will also improve the output efficiency of the input. In the residual value A, there is not only the contribution of generalized technological progress, but also, more importantly, the contribution of market-oriented reform to improve the production rate of factors. Having a very long period of transition, China is a dualistic system country with the largest population, who has a land area that ranks in the top in the world and becomes the second largest economy. When observing the cyclic fluctuation of economic growth of China, it is found that the average annual growth rate of GDP between 1981 and 1985, 1991 and 1994, 2001 and 2005 are 10.17%, 11.67% and 9.78% respectively, while the average annual growth of TFP in these periods are 6.53% [19], 6.65% [20] and 3.30% [21] respectively. The data of the growth of TFP and GDP in these three periods shows an inverted U shape [22]. However, no prompt arrival of any revolutionary technological progress in a broad sense in China during these three periods has been seen, instead, they are all the periods of the start and upsurge of vigorous institutional reforms and opening up. Our decomposing of the average annual TFP growth of 3.10% since China's reform and opening up, shows that only an approximate 1.00% of the growth is realized by technological progress in a broad sense, and 2.10% of it is realized by improving the utilization efficiency and allocation of factors through the institutional reforms.

Generally speaking, the developed countries of market economy do not have the same distortion in dualistic system as the countries in the process of transition, and there is no TFP formed by the system transition countries' vigorous reform to improve factor productivity and its allocation. The main source of TFP is technological progress in a broad sense. The conclusion of some empirical research tells that the TFP growth curve of the developed countries is a stable curve that almost parallels with the X-axis in the long term, although there are some small fluctuations. The empirical study of Jones [23] found that in the United States and Europe from 1980 to 2000, the growth of TFP was flat while the R&D personnel expanded continuously (with a right-leaning or even exponential growth). From 1970 to 2012, the average annual growth rate of TFP in the United States, Europe and Japan was only 0.90%, 1.00% and 0.70% respectively, and Korea, as a newcome developed country, was only 1.60%. Fispage [24] believes that the balanced growth theory simplifies the analysis of the process of technological progress, which has an almost linear, clear and constant relationship among R&D investment, technological progress and economic growth. Although the TFP growth trend caused by technological progress in a broad sense has small fluctuations, yet it is a flat curve.

There is a reason for that. In the early stage of industrialization and before that, humans' domesticating animals which are used in agriculture, planting looms and agriculture, and the invention and application of technology like electric power and steam-powered spinning wheels, etc., can stimulate the explosive growth of specialized animal husbandry and agriculture with a large scale, and can also be helpful to produce products by mechanical power, in a way of specialization, standardization and scale production, which can make the wealth grow in a more explosive way. When reaching the middle and later stages of industrialization, especially in the post-industrial society, we encounter a situation in which almost no more technologies for production of the large-scale material wealth can be explored, and technological innovation has turned its direction to the information networks and digital intelligence. Such kind of technologies can both generate new growth and somehow "save" the value added in wages, interest and rent by displacing human and physical costs; Moreover, technological progress in the middle and late stages of industrialization has become a comprehensive and composite innovation process with crossovers, and it is increasingly difficult to form a J-shaped growth of TFP for every technological innovation alone.

As is shown in Figure 2, in the period of vigorous reforms, the TFP growth curve formed by the reforms comes into a shape of an inverted U-shaped because the institutional transition improves the allocation of idle factors and the output rate of factors which are under a low utilization, while the TFP growth curve under the technological progress in a broad sense is almost a curve which nearly parallels with the X-axis. This finding should be a key to understand and reasonably explain the TFP source of China as a country in the institutional transition of the dualistic economic system from a methodological perspective-- we assume that the output contribution growth of the technological progress in a broad sense is a flat curve that parallels with the X-axis in both developed countries with market economy and the countries in the process of institutional transition, so that the integral between the inverted V-shaped or inverted U-shaped TFP growth curve formed in China's previous institutional reform cycle and the flat curve, should be the TFP output from the improvement of factor utilization rate and the efficiency of allocation brought about by the institutional reforms.

It should be pointed out in particular that the estimation of TFP growth brought by China's previous institutional reforms of the dualistic economy is an estimation of the "residual value" without such an analysis logic as "the institutional distortion of factor allocation in the dualistic system - irremediable distortion - idle factor and low utilization - institutional factor slack". In other words, methodologically, we can include the part of the added value of the output when estimating the TFP.

$$Y = \frac{\Delta A_1}{A_1} + \frac{\Delta A_2}{A_2} + a\frac{\Delta W}{W} + b\frac{\Delta K}{K} \qquad (1)$$

In the model(1), Y stands for GDP growth rate; W for Labor force; K for Capital; $\Delta A_1$ for general technological progress TFP; $\Delta A_2$ for system reform factor allocation improvement TFP, （a+b）=1.

From 1977 to 2018, the average annual growth rate of TFP in Japan and South Korea was 0.8% and 1.03%, respectively. Although the fluctuation range was slightly larger than that of the United States and Europe, it was also an almost parallel curve. In China with the dualistic system of the same period, its TFP growth rate was an average of 3.43% per year, and during the period of greater reform and opening up in 1981-1985, 1991-1995, and 2001-2009, the upward fluctuation range was greater. Then, assuming that 1% TFP growth in capitalized countries is the standard value, and China's most optimistic estimate for catching up with those country is 1.2%, then the entire TFP growth curve of China can be subtracted to obtain the TFP growth part of the reform. from 1978 to in 2018, the average was 2.23%.

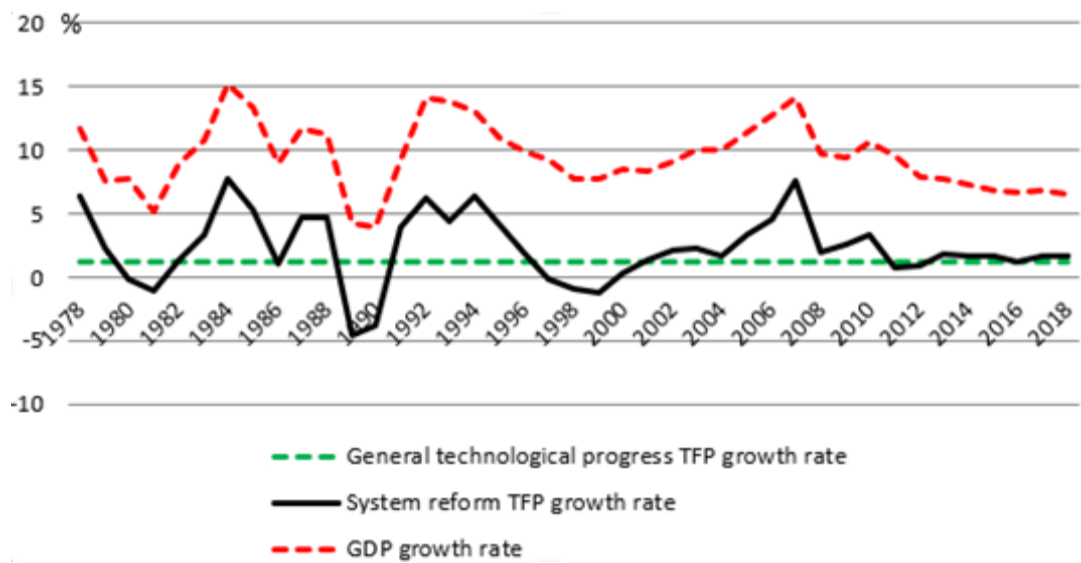

Figure 2: The relationship between China's reforms TFP growth and general technological progress TFP growth

(Data source: National Bureau of Statistics website database-data.stats.gov.cn/easyquery.htm?cn=C01; Data source: Asian Productivity Organization (Asian Productivity Organization); Website address: http://www.apo-tokyo- aepm.org.)

However, in the scenario of marketization, it can find out the price of idle and lowly-utilized factors, the elasticity of substitution between factors and other parameters. Theoretically, we can see it in this way-before the reform of marketization, the input factors that are actually contributing to the output are less than the nominal input factors due to the distortion of the system; thus, through the reform of marketization on factors, the idle and low utilization of the nominal input factors should be eliminated, and their allocation and utilization rate should be improved, so that the input factors that are actually contributing to the output can have a smaller gap with the nominal input factors. In this way, we can directly put the added output part formed by vitalizing the stock onto the numerator in the function of the input-output and calculate the potential of economic growth more accurately in an endogenous approach.

**Land Increase and Allocation Reform: A Regression of the Model of Economic**

**Growth from Neoclassical to Classical**

Marx [25] quoted "labor is the father of wealth, land is the mother of wealth" from William Petty when describing the necessary conditions of production. On this basis, Adam Smith [26] also proposed that "all products of a country can be decomposed into three parts: land rent, labor wage and capital profit". But, as mentioned earlier, neoclassical economists excluded land from their models of economic growth. However, in view of the particularity of China's long-term transition of the dualistic system, the reform and development of land factor allocation to increase land utilization rate are of great significance for seeking new growth potential of China. The basic mathematical expression that explains the coupling of this land ownership structure and economic operation as well as growth has two aspects as follows.

Firstly, the independent variables of the factor of land and the dependent variables of land rent contribution of the countries in the process of transition should be added into the input-output model, that is to say, it is a regression from the model of neoclassical economy where has the equation of "output = residual value + labor contribution + capital contribution" to the model of classical economy whose equation is "output = residual value + labor contribution + capital contribution + land contribution" instead. The reasons why we have to make this regression area follows: (1)In the countries of market economy, land can be transacted, can be priced, has already been capitalized, monetized and has gone through the process of marketization; however, a large part of China's land is not transacted in the primary market, and almost all land cannot be transacted in the secondary market, and there is no capitalization, monetization and marketization. (2) In terms of the reform process of the allocation system of the three major factors in China, the market-oriented reform of the capital factor comes first, the market-oriented reform of the labor factor is right in progress, and the market-oriented reform of the land factor has just begun its way due to the planned adjustment of factor allocation, planning administration and controls of the use which have a profound and deep-rooted influence. (3) As mentioned above, a large amount of urban and rural land is prohibited from transacting, or though allowed to trade yet still blocked by the dualistic system, which results in a large amount of idle and lowly-utilized land, and that is what is called the "institutional land slack". When calculating the relationship between the reform and economic growth in the country with a dualistic system, it is necessary to calculate the new contribution of the output from the reuse of institutional land slack after the reforms, so as to estimate the economic growth potential brought by the market-oriented reform on the factors of land more accurately. (4) As a country of the dualistic system whose degree of the development of land is still low, China's land utilization is only 71%①[3], which means there is much room for the enhancement of utilization level by adjusting the allocation of water resources as well as the south-north water diversion project, transforming and exploiting the unused land, and increasing the amount of available land, which makes the curve of land supply change

---

[3] It is calculated according to the "Statistic Report of the Land, Mineral and Marine Resources of China, 2017", and all data can be obtained in http://www. mnr. gov. cn /sj /tjgb /

into a right-leaning curve from a vertical one. (5) In the countries of market economy, due to land's being capitalized, its appreciation will not be large; however, China's huge amount of rural land and urban allocated land, has not been traded in the past, or their transaction is prohibited, so that their market value turns out to be zero basically. Once the reform is carried out to make it able to be transacted and input, its value will rise from zero to the market price.

Secondly, in the actual national economic accounts of countries of modern market economy, here are some rules as follows. (1) Rents of land and housing are included in GDP. All the rents including the rents paid for place-using, cultivated lands, factories and dwelling generated through the property companies, agricultural landowners, rural collectives, and urban residents' renting their houses and land, have to be added into GDP. (2) In the fixed assets where land and buildings cannot be separated, they are included in GDP in the form of depreciation and profit. Land and buildings are inseparable, and in the accounting standards, the contribution of land to the output is included in the contribution of capital profit. (3)According to the common international rules of the national economic accounting, one can be considered to pay the housing rent to oneself even he or she owns or lives in the house according to the Wieser's attribution theory of factor productivity contribution, and an estimated rent need to be added into the GDP after the subtraction between the estimated data which the statistical bureau has made according to the renting level as well as the national housing area and the actual data of the rented house[27].

So far, according to our previous discussion, we can build up a basic model which can be foreseen and calculated under the growth theories of classical economy, can depict the history of the reforms on the dualistic system and the economic growth, and can also forecast the future transition from the dualistic system of factor allocation where the command economy and market economy coexist to the unary socialist market allocation economic system and its future economic growth relationships.

$$Y = \frac{\Delta A}{A} + a\frac{\Delta W_f + \Delta W_{rn}}{W} + b\frac{\Delta K_f + \Delta K_{rn}}{K} + c\frac{\Delta L_f + \Delta L_{rn} + \Delta L_{in}}{L} \qquad (2)$$

In the model (2), Y denotes the growth rate of GDP; A stands for technological development level, W for labor force, K for capital, L for land; F represents the input and output increment of the original factors, "Rn" represents the output increment of the institutional factor slack after the reform of the marketization, and "in" stands for the input and output increment of new land. The "a" stands for the elasticity coefficient of labor output, "b" for the elasticity coefficient of capital output, and "c" for the elasticity coefficient of land output, and their relation appears to be "a + b + c = 1".

Based on the above model (2), the potential economic growth rate of market-oriented reform on factors, and the average annual GDP growth rate from 2021 to 2035 are estimated to be in a range between 5.25% and 6.30%. Formed by three aspects concerning the reform of allocation of crucial factors, the extension of the land development strategies, and the promotion of innovation in the future where each one has a relatively high and a relatively low target of reforming plans, the dynamic growth patterns are as follows:

(1) The annual growth rate brought about by the input of factor, vitalizing the major factors and new added factors of land is 3.25% or 3.80%, which are respectively low and high target. (2) An annual growth rate of approximate 1.00% can be brought about by the increment of value, market transaction, asset financing, the wealth effect and property income which are formed by capitalization reform of the homestead and other land for construction. (3) The TFP growth by promoting technological progress in a broad sense and the innovation also has a low and a high target, which can be an annual average of 1.00% or 1.50% respectively [12].

## E. The Economic Equilibrium and Economic Growth in the Country in the Long-term Transition Process of Dualistic system

Through the analysis above, the mathematical relationship between the market-oriented reform of production factors and economic growth from the output and supply side has already been analyzed. However, on one hand, economic growth also needs the balance between aggregate supply and aggregate demand, otherwise an insufficient aggregate demand will lead to the overproduction; on the other hand, from the perspective of price level, debt chain, currency value and stability of the financial system, it is also necessary to take into consideration the smooth and safe operation of the national economy and economic growth.

**Economic System Distortion and the Imbalance of the chain of "Distribution - Consumption - Production" as well as the Rectification on It**

Economic growth on the supply side needs to be balanced according to the market capacity on the demand side. If the distortion in the circulation process of "production -- allocation -- expending -- consumption -- production······" is not effectively contained, there will not be a sustainable development and the capacity of the high-speed growth in productivity on the demand side even if the supply side has the strong production capacity formed by the intelligent manufacturing and the wealth of the digital economy and so on in the future. Under such circumstance, China's economic growth total supply and total demand imbalance problem can be very difficult to analyze, describe, and balanced growth just by the medium or short-term equilibrium growth model.

Not only does the dualistic economic system distort the process of the economic system of "production - distribution - consumption", but also it distorts the reproduction of population and labor force, which results in a long-term iterative contraction of the process of "employment - income - consumption". Also, such distortion not merely distorts the optimal allocation of factors, but also distorts the income distribution of residents and further inhibit residents' spending ability as well as distorting the balance between demand and consumption, which normalizes the excess of production capacity. In terms of the distortion of China's dualistic economic system in a particular long period and the transformation of its development strategy, as shown in Figure 3, after the entry of the 21$^{st}$ century, what the demand of China has seen first comes to the relatively shortage of the consumption demand compared to the supply, which is caused by the relative and absolute iterative contraction of population and labor force, and then comes the relative excess of the domestic production capacity caused by the decline of the ratio of the export in GDP, after which is the transmission and crowding-out of the income and consumption of rural and urban residents due to the land fiscal arrangement and high housing price.

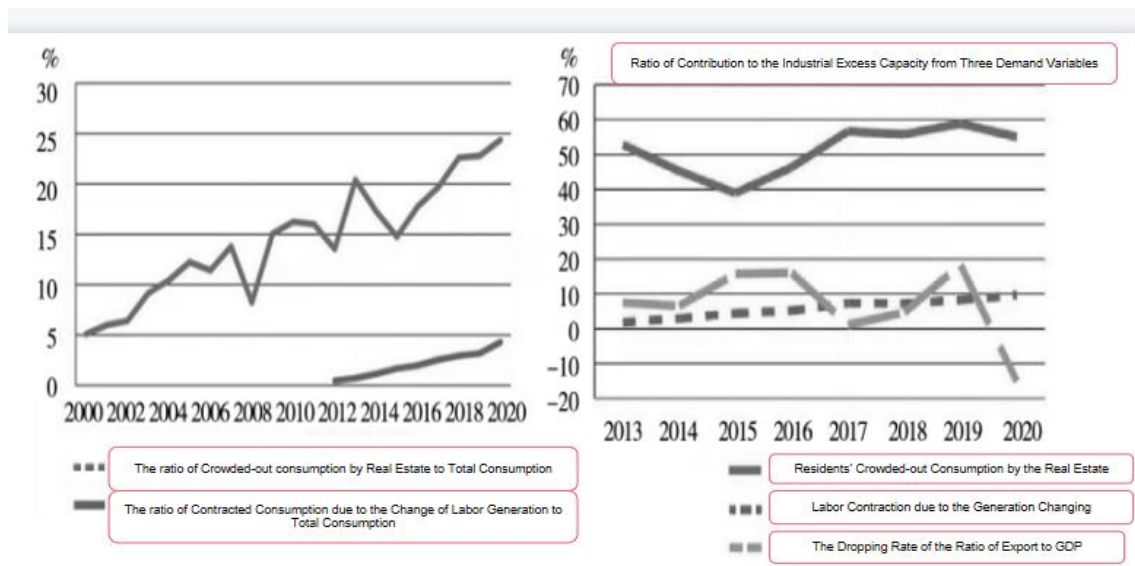

(Figure 3: The Transferring and Crowding-out Effect of Land Finance and High Housing Price on Residents' Income, and Industrial Excess Capacity due to Relative Contraction of Three Kinds of Demand)

First, the iterative contraction of population and labor force leads to the iterative contraction along the chain of "employment - income - expenditure - consumption". From the perspective of the transition of China's population growth, the TFR dropped to the iterative equilibrium point of 2.10 in the early 1990s during the early or middle stage of industrialization and urbanization. Since then, it has dropped rapidly and hit around 1.20 in 2020. The reason is that since the reform and opening up, the direct and opportunity costs are getting increasingly higher due to the birth control policies with excessive forces and an extremely long time, and market-oriented birth and upbringing. In particular, since the mid-1990s, an inertia of low fertility which resulted in the imbalance of population reproduction, has been formed because of the culture inclining to a low birth rate, urbanization of the population, the improvement of the women's level of education involvement and participation in the work force, as well as the high cost of education, health care and housing. That is to say, the population of the next generation is less than that of the previous generation. If the fertility rate is less than 2.10 twenty years ago, and when the rate of population increase goes downwards, the internal interaction process between the growth of population the total supply and demand of national economy appears to be like this - 20 years later, growth of the new population getting employed continues going downward, and with other factors being abstracted, we can see a negative growth of the new population getting into their jobs, which makes next year's total demand continuously less than this year's supply capacity. From the perspective of iterative decrease, the main population in the economic system has shrunk to 3.45 million in 2012 and will shrink to 39.83 million in 2020. The scale of household consumption has shrunk to 766.34 billion RMB in 2012, and by 2020, the cumulative scale will have shrunk to 12774.41 billion RMB. Initially, though, the contraction of household consumption in GDP takes up a ratio of 0.15% in 2012, which seems weakly significant, yet this ratio will increase to 1.25% in 2020 due to the iterative cumulative contraction of the main population in the economy.

Secondly, China's export-oriented industrialization tends to cease to go on, and the driving force of export demand for economic growth declines, due to the rising domestic labor costs, the rise of international trade protectionism, the repatriation of overseas manufacturing led by some developed countries to compete with the export from developing countries. Seeing from the stage of industrialization, in 1978, especially during the integration into the international cycle in the 1980s, China has transformed from an economic development model based on satisfying domestic demand to a model driven by the export demand. In 2000, China became a country in the stage typical export-oriented industrialization and this stage lasted 16 years. Since 2016, China has shifted from a country with an export-oriented industrialization model to a country with a domestic-demand-oriented model. In 1978, China's exports came to a total of 16.8 billion RMB, accounting for only 4.56% of its GDP at that time. If the 20% and above of exports in GDP of the developing and the emerging industrialization countries is regarded as the dividing line of export-oriented development model, then China has been above this line from 2000 to 2015, during which the total export in 2006 peaked at 7.7597 billion RMB which accounts for 35.36% of GDP. In 2020, the total export reached 17.9326 billion RMB, which accounted for 17.65% of GDP.

Finally, as mentioned above, the land finance of local governments takes away farmers' property transaction income with an excessively high proportion of distribution, and the high price of houses supplied through single development and construction channel of the real estate transfers the income of urban residents who purchase houses, and crowds out the consumption expenditure ability of urban and rural residents. As can be seen from Figure 3, the excess of production capacity caused by such situation is much higher than that caused by the iterative contraction of population and the decline of export demand. Land finance and high housing prices distort the economic process of "production - distribution - exchange - consumption", resulting in a serious shortage of consumption demand of residents. For example, in 2020, local governments receive 8.414.2 billion RMB in land transfer fees, and assuming that 50% of that is supposed to be left for farmers, the rural residents' income which will be transferred will reach 4207.1 billion RMB, which thus means a depriving of 33675.7 billion RMB from peasants' consumption and expenditure capacity. The high housing price provided by a single channel of real estate also squeezes the other consumption expenditure capacity of 3645.8 billion RMB from urban residents. That year, the industrial output excess was 13.15 trillion RMB, which includes an excess of consumer goods of approximate 6.58 trillion RMB. Thus, it can be clearly seen that without land finance and high housing prices, China would not suffer from industrial excess capacity.

Transaction of rights of property use of the rural land as well as the entrepreneurship based on factor of land is the key to the increase of farmers' income. In the next 15 years, from the perspective of the changes in the various types of rural income, the government will allow 1.50% of the assets of rural land to be transacted every year, and a present discounted value of the annual property income of 4.5 trillion RMB is likely to be generated for the peasants even if the government collects 30% of the tax. If peasants are allowed to start businesses based on factor of land and to utilize 20% of rural homestead and other construction land in the secondary and tertiary industries, there will also be an annual income of 5 trillion yuan from the input of the factor of land. Annually, there will be a total of consumption demand of 6.65 trillion yuan.

Marketization of land and multi-channel housing tendency can reduce relative and absolute housing prices and release urban residents' consumption. The market-oriented transaction of homestead, as well as the two-way replacement of urban and rural population, labor force and homestead will be permitted, cooperative house construction in cities can be opened up and the supply of public rental housing will increase. Gradually the price-to-income ratio can be reduced from around 9.30 to around 6.00, and thus restoring the urban residents' annual spending capacity of 3 to 4 trillion yuan, which has been squeezed by high housing prices and mortgage repayment. The National Economic Engineering Laboratory of Dongbei University of Finance and Economics has simulated the related flow, and the rectification of "income - output - consumption" direction and flow by the above-mentioned market-oriented reform can provide the industrial production with about 10 trillion yuan of new consumption demand capacity every year, which can effectively reduce excess capacity, stabilize and strengthen China's manufacturing industry, and help assure the medium of high speed of China's future economic growth on the demand side.

**Debt-type Economic Model and The Security of Operation and Growth of National Economy**

Economic growth in the field of production requires not only an achievable capacity of market demand, but also stability and security of the debt chain and the value of the currency. There should be no systematic breaking in the chain of debt, large currency fluctuations or sharp drops in the exchange rate, which can result in a vertical drop in per capita GDP.

As mentioned above, as for the debt and related money supply of market economy countries, assets of the land and housing are the basis of their credit and currency value, which is an indispensable guaranteed mechanism. The ratio of China's total non-financial debt takes up from 101.70% in 1995 to 272.00% in the GDP in 2020. The debt stock is of a bad quality, and the government will be suffering from great pressure to service the principal and interest in the future, as well as the huge gap between medium or long-term pension income and expenditure. The ratio of supply of M2 in GDP has risen from 99.00% to 217,00%, a level much higher than that of many of the world's major economies. This shows that the traditional monetary theory of QP = MV cannot explain the relationship between GDP, monetary scale, monetary velocity and price anymore.

Over the next 15 years, the percentage that debt and the currency supply take up in GDP will undoubtedly continue to rise because of an aging population structure and the need to support a sluggish economy. From the perspective of the three input factors of labor force, capital and land, labor force cannot become the guarantee of credit of the debt economy in the future. Capital used for debt mortgage except for the fixed assets is likely to increase more economic bubbles, and thus the best foundation to stabilize the debt chain and currency value is the real estate such as land. However, in China, except for approximately 35 billion square meters of urban housing which can be transacted or mortgaged, there is still no secondary market for all kinds of land and homestead in rural areas that is worth about approximately 500 trillion yuan, and over 300 trillion yuan of land in urban and county areas. Large-scale land and homestead in urban and rural areas cannot be traded or mortgaged so that there is no anchor of credit guarantees and currency values to stabilize an economy with high debt and multiple currencies. With these constraints, we propose the following model for our analysis.

The stock, price and debt balance of land and buildings assets are used as variables and incorporated into Fisher's model. The supply side is GDP, flow assets and stock assets. The substitution relationship between GDP and assets to the money supply needs can be observed. The demand side is the debt balance, establish the mathematical relationship between debt balance and the money supply. In this way we can observe and try to understand the formation mechanisms of the aforementioned Fisher's paradox in the debt expansion and accumulation economy. In this regard, we have constructed tree models for basic analysis:

Definition of the supply side model:

$$a \times (GDP \times P_q) + b \times (A_1 \times P_{a1} + A_2 \times P_{a2}) = M \times V_s; \qquad (3)$$

Definition of the demand side model:

$$M \times V_d = D \times P_r; \qquad (4)$$

Definition of debt to GDP and assets balance：

$$a \times (GDP \times P_q) + b \times (A_1 \times P_{a1} + A_2 \times P_{a2}) = D \times P_r \times V_f \qquad (5)$$

Pq stands for GDP nominal price increase rate, A1 denotes the flow land (used for construction of residential and commercial buildings cannot be re-traded) asset, Pa1 for flow building land asset price, A2 for stock land and houses (Land and houses can continue to be traded) Assets, Pa2 stands for stock price of urban land and houses, a is the proportion of GDP in total supply-side factors, b is the proportion of assets in total supply-side factors, (a+b)=1, Vs for the currency relative to the supply-side GDP and stock asset balance turnover speed, D for debt, Pr is the debt interest rate, Vd for the currency relative to the turnover speed of debt, Vf for the debt relative to the supply-side GDP and stock asset balance turnover speed.

If GDP is optimistically estimated to grow at an average annual rate of 6.10% in the future, and the debt balance and currency supply are conservatively estimated to grow at an average annual rate of 10.00% and 8.00% respectively, that is to say, the average annual increment of these three parts will be 6.80 trillion yuan, 57.60 trillion yuan and 31.70 trillion yuan respectively. By 2035, the ratio of debt balance to GDP will rise to 465.00%, and the ratio of M2 in GDP will rise to 327.00%. The national Economic Engineering Laboratory of Dongbei University of Finance and Economics simulated the situation of deepening of national economic debt under four different conditions of: being not capitalized, being capitalized for a small part, being half capitalized and being totally capitalized. Under the first condition shows the earliest collapse of financial and economic system. Then the upcoming crash is the scenario of the second condition; the third condition can be maintained for a while, but debt continues to grow excessively and will eventually collapse.

If land, homestead, and other newly increased available land for water transfer and land improvement are undercapitalized reform, at present, the assets of all kinds of land that can be transacted and mortgaged in rural areas and the land that can be transacted in urban industrial and mining enterprises are calculated to be about 500 trillion yuan and 150 trillion yuan respectively based on shadow prices. In the future, if China implements the strategy of adjusting the distribution of water resources, transforming the unused land and increasing the amount of available land, and makes the new land being able to be transacted and mortgaged, there will be an addition of the land assets of 145 trillion yuan. Also, in the future, China will have 800 trillion yuan of stock assets which can be used as the credit guarantee of debt and currency and the stable anchor of currency value. On average, 6.70% of the stock land that can be transacted and mortgaged will be released as flowing assets annually. With the average annual increase of the flow assets of the available land thanks to the strategy of water transferring and land improvement, there will be 53.30 trillion yuan of assets available for exchanging and mortgaging annually, and 6.80 trillion yuan of new-increased GDP annually in the future. All these could well serve as an anchor for currency security and credit guarantee of the currency supply as well as debt expansion.

## F. Conclusion

In such a transitional country as China, the coexistence as well as the conflict of the dualistic system will be inevitable, and a variety of risks and challenges will have to be faced with. We should think rationally, make reasonable decisions based on scientific analysis and take actions. More importantly, we should find out the crux of the issues and at the same time to seek for solutions through the analysis of economic theory. In the next 15 years, it is necessary to push forward the reforms including establishing an appropriate level of macro tax burdens, forming a modern state-owned enterprise system, liberalizing household registration, capitalizing the factors of land and executing the market-oriented allocation of factors of production, to implement the strategy of water transferring and land improvement, to change the unfavorable factors into the favorable ones, and thus to find new potential for economic growth.

There are four advantages that could be found in China's future development and growth: (1)the post-reform advantages from the changing of the dualistic system of command and market economy to the socialist unary market economy system, which release huge productivity; (2)the advantages of new development and rising as a large developing country with a low level of development, which will focus on adjusting the distribution of water resources, increasing the available land and improving the utilization rate of land; (3)another developmental advantage of the internal virtuous cycle of production together with the demand and the balanced growth of national economy, since nearly 800 million Chinese rural residents are still in the middle stage of industrialization and enjoying the relevant level of economic development welfare, whose income level and consumption ability will be enhanced in the future; (4) the incomparable security advantages from the large-scale land capitalization reform that is provided for the smooth macro-control and operation of China's national economy, which is second to none in the world.

From the analysis of this paper, China's economic system reform has reached the threshold of the transformation reform from dualistic system of command and market economy to the socialist unary market economy system. It is estimated herein that the market-oriented reform of land factor allocation will be launched since 2021, together with the implement of the strategies water transferring and land improvement, as well as further and tough reforms on the state-owned enterprise system. In this way, through the vigorous reform spanning for 7-8 years, new growth potential in all aspects will be obtained. If the national economy can achieve stable and secure growth around 6%, 5%, 4% respectively in the next three periods of five years, then there is no doubt that China will become a country of preliminary modernization as well as one of the high-income countries.